\documentclass[
    prd, twocolumn, superscriptaddress, nofootinbib, amsmath, amssymb,
    aps, floatfix, preprintnumbers, longbibliography
]{revtex4-2}
\usepackage[utf8]{inputenc}
\usepackage{times}
\usepackage{setspace}
\usepackage[dvips]{graphicx}
\usepackage[dvipsnames]{xcolor}
\definecolor{CiteBlue}{RGB}{45,52,151}
\usepackage[
    colorlinks=true,
    linkcolor=CiteBlue,
    urlcolor=CiteBlue,
    citecolor=CiteBlue
]{hyperref}
\usepackage{aas_macros}

\usepackage[capitalise]{cleveref}
\crefname{sec}{Section}{Sections}

\usepackage{siunitx}
\DeclareSIUnit{\year}{yr}
\DeclareSIUnit{\au}{au}
\usepackage[normalem]{ulem}
\newcommand{\refcite}[1]{Ref.~\cite{#1}}
\newcommand{\refscite}[1]{Refs.~\cite{#1}}

\usepackage{bm}
\newcommand{\bb}[1]{\bm{\mathrm{#1}}}
\newcommand{\du}{\mathrm d}
\newcommand{\dd}{\,\du}
\newcommand{\pbh}{\mathrm{PBH}}
\newcommand{\sso}{\mathrm{SSO}}
\newcommand{\fom}{q_{\mathrm{fom}}}
\newcommand{\fommax}{{\fom^{\mathrm{max}}}}
\newcommand{\fommin}{{\fom^{\mathrm{min}}}}

\begin{document}

\preprint{MIT-CTP/5650} 

\title{Close encounters of the primordial kind:\protect\\a new observable for primordial black holes as dark matter}

\author{Tung X. Tran}
\email{tungtran@mit.edu}

\affiliation{\ignorespaces
    Center for Theoretical Physics,
    Massachusetts Institute of Technology,
    Cambridge, MA 02139, USA
}

\author{Sarah R. Geller}
\email{sageller@ucsc.edu}

\affiliation{\ignorespaces
    Center for Theoretical Physics,
    Massachusetts Institute of Technology,
    Cambridge, MA 02139, USA
}
\affiliation{\ignorespaces
    Santa Cruz Institute for Particle Physics,
    Santa Cruz, CA 95064, USA
}
\affiliation{\ignorespaces
    Department of Physics, University of California, Santa Cruz,
    Santa Cruz, CA 95064, USA
}

\author{Benjamin V. Lehmann}
\email{benvlehmann@gmail.com}

\author{David I. Kaiser}
\email{dikaiser@mit.edu}

\affiliation{\ignorespaces
    Center for Theoretical Physics,
    Massachusetts Institute of Technology,
    Cambridge, MA 02139, USA
}

\date{\today}

\begin{abstract}\ignorespaces
    Primordial black holes (PBHs) remain a viable dark matter candidate in the asteroid-mass range. We point out that in this scenario, the PBH abundance would be large enough for at least one object to cross through the inner Solar System per decade. Since Solar System ephemerides are modeled and measured to extremely high precision, such close encounters could produce detectable perturbations to orbital trajectories with characteristic features. We evaluate this possibility with a suite of simple Solar System simulations, and we argue that the abundance of asteroid-mass PBHs can plausibly be probed by existing and near-future data.
\end{abstract}

\maketitle

\section{Introduction}
\label{sec:introduction}

Our Universe contains a substantial amount of dark matter (DM), per decades-old consensus \cite{deSwart:2017heh,Peebles:2022bya}. Despite well-motivated theoretical models for various particle DM candidates spanning nearly fifty orders of magnitude in mass (see e.g. \refscite{Boveia:2018yeb,Weltman:2018zrl,Agrawal:2021dbo,Cooley:2022ufh,ParticleDataGroup:2022pth}), there is as yet no direct evidence that DM consists of a new elementary particle species. In recent years, an alternative hypothesis has regained traction: that perhaps much or all of the DM consists of primordial black holes (PBHs) \cite{Zeldovich:1967lct,Hawking:1971ei,Carr:1974nx}. Numerous observables restrict the properties of PBHs that could account for DM, but a window of about six orders of magnitude in mass remains fully unconstrained. For masses $\qty{e17}{\gram} \lesssim M_\pbh \lesssim \qty{e23}{\gram}$---that is, for masses typical of asteroids---PBHs could account for the entire DM abundance. (For recent reviews, see \refscite{Khlopov:2008qy,Carr:2020xqk,Green:2020jor,Villanueva-Domingo:2021spv,Escriva:2022duf}.) 

The constraints at the boundaries of this region are well established. The upper limit is set by constraints from extragalactic microlensing surveys \cite{Sugiyama:2019dgt,Smyth:2019whb}. The lower limit arises from efficient Hawking evaporation, since high-energy cosmic rays emitted by light black holes would be readily detectable \cite{Belotsky:2014twa,Carr:2016hva,Boudaud:2018hqb,Laha:2019ssq,DeRocco:2019fjq,Laha:2020ivk,Chan:2020zry}. (See also \refscite{Khlopov:2008qy,Carr:2020xqk,Green:2020jor,Villanueva-Domingo:2021spv,Escriva:2022duf}.) However, the asteroid-mass range has proven difficult to probe. Originally, microlensing constraints were thought to apply in this regime as well as at higher masses \cite{Niikura:2017zjd}. However, \refscite{Sugiyama:2019dgt,Montero-Camacho:2019jte,Smyth:2019whb} demonstrated that the effects of wave optics and finite source sizes severely weaken lensing constraints below \qty{e23}{\gram}.\footnote{These effects were later incorporated into an updated version of \refcite{Niikura:2017zjd}.} Moreover, limits from stellar capture estimated by \refcite{Capela:2013yf} were later weakened by updated arguments in \refcite{Montero-Camacho:2019jte}. As a result, PBH DM remains largely unconstrained in the asteroid-mass regime, and the future of the PBH DM hypothesis depends crucially on probes of this portion of the parameter space.

In this paper, we consider a novel means to probe PBHs in the asteroid-mass range. We focus on the effects of PBHs on the motions of visible objects, similarly to \refscite{Dror:2019twh,Li:2022oqo,Belbruno:2022hsm,BrownHeUnwin,Bertrand:2023zkl,Cuadrat-Grzybowski:2024uph}. In particular, in order to achieve the highest possible sensitivity, we propose the use of the highest-precision astronomical measurements available: Solar System ephemerides. Given more than five decades of lunar laser-ranging data \cite{Murphy:2013qya,Battat:2023upl,Colmenares:2023jlw}, more than two decades of precision monitoring of Mars orbiters \cite{Konopliv2016,Yan2018}, and sophisticated tracking of the motions of thousands of smaller objects \cite{JPLeph,ParkJPL2021,Pariseph,INPOP21a,Fienga:2023ocw}, our own Solar System is richly instrumented for the detection of massive interlopers. We therefore evaluate the observable effects of low-mass PBH encounters on the revolutions of the celestial spheres \cite{Copernicus:1543}.

In particular, we study the prospects for detecting perturbations to the measured distances between Earth and Solar System objects (SSOs) due to PBH flybys. If they account for all of DM, the abundance and phase space distribution of PBHs can be readily determined---and if $M_\pbh$ lies in the asteroid-mass range, then at least several flybys through the Solar System should be expected to have occurred over the lifetime of modern observing programs. The expected rate of observable events depends both on the PBH mass distribution and on the precision with which the motions of various Solar System objects are presently monitored. As we demonstrate here, if the entire DM abundance consists of PBHs with masses $\qty{e18}{\gram}\lesssim M_\pbh \lesssim \qty{e23}{\gram}$, then Solar System ephemerides could plausibly detect PBH flybys at a rate of order once per decade. This could either enable a quasi-direct detection of PBHs or place new constraints in the asteroid-mass range.

This paper is organized as follows. In \cref{sec:encounters}, we perform a simple order-of-magnitude estimate of the expected effects of a PBH transit on Solar System observables, demonstrating a favorable detection rate for perturbations to the distances between Earth and Solar System objects from flybys at an impact parameter of several \qty{}{au}. In \cref{sec:simulations}, we improve on our analytical estimate with simplified Solar System simulations, and we identify the expected properties of observable signals, focusing on the distances between Earth and various inner planets (Mercury, Venus, and Mars). We justify this choice with a discussion of damping mechanisms that might attenuate a signal of a PBH transit, and argue that our minimalist simulations are representative of the qualitative features of transits. In \cref{sec:detection}, we analyze ensembles of simulated PBH transits, and estimate the sensitivity of Solar System ephemerides to transits with varying parameters. \Cref{sec:discussion} summarizes our proposal for a new means of probing asteroid-mass PBHs. We also discuss next steps, particularly the incorporation of more sophisticated numerical methods that include additional physical effects. We highlight the prospects for our proposed observable to test the hypothesis that PBHs account for all of DM.

Throughout this paper, we work in natural units with $c = \hbar = k_{\mathrm{B}} = 1$ unless otherwise noted.

\section{Encounters in the Solar System}
\label{sec:encounters}

In this section, we introduce our proposed observable, and perform a rough estimate of the size of the effect. This leads us to a first estimate of the rate of detectable flyby events. We then discuss the physical effects that complicate a proper computation of the flyby signature, motivating the application of numerical computations in the following section.

\subsection{Detecting PBH flybys}
\label{sec:flybys}
If a PBH were to transit our Solar System, it would perturb the motion of visible objects. State-of-the-art simulations of the motions of objects within our Solar System include the DE441 model maintained by the Jet Propulsion Laboratory \cite{ParkJPL2021} and the INPOP21a model maintained by the Observatoire de Paris \cite{INPOP21a}. (See also \refscite{Vallisneri_2020,Grimm_2022}.) The JPL DE441 model, for example, incorporates the Sun and eight planets, nearly 300 planetary satellites, more than 1.3 million asteroids, and almost 4000 comets. 
Such simulations depend on high-precision observations of various objects for benchmarking. Among the most precise data used in these ephemerides concern the Earth--Moon distance and the Earth--Mars distance. For the Earth--Moon distance, lunar laser ranging has succeeded in achieving an accuracy of $\mathcal O(\qty{1}{\milli\meter})$ over the past fifteen years \cite{Murphy:2013qya,Battat:2023upl}, while precision monitoring of various Mars orbiters and landers since the late 1990s yields an accuracy of $\mathcal O(\qty{10}{\centi\meter})$ for the Earth--Mars distance \cite{Konopliv2016,Yan2018,INPOP21a}.

Motivated by the extraordinary precision of these measurements, we consider the Solar System as a single large \emph{compact object detector}. The calculation of the rate of observable flybys is similar to the calculation of the event rate in a DM direct detection experiment. For simplicity, we assume a monochromatic mass distribution for PBHs at a mass $M_\pbh$. Given the mass density $\rho_\pbh$, the number density is $n_\pbh = \rho_\pbh / M_\pbh$. If PBHs account for all of DM, then $\rho_\pbh \simeq \qty{0.4}{\giga\electronvolt/\centi\meter^3}$ in the solar neighborhood \cite{Pitjev,BinneyWong2017,PostiHelmi2018}. In particular, within a sphere centered on the Sun with radius equal to the orbit of Jupiter (\qty{5.2}{\au}), one expects to find a number of PBHs given by
    $N_\pbh = 1.4\times[M_\pbh/(\qty{e18}{\gram})]^{-1}$
at any given time.
Moreover, these objects should have a Maxwellian distribution of velocities, with a typical speed of $v_0 \approx \qty{220}{\kilo\meter/\second}$ \cite{BinneyWong2017,PostiHelmi2018}. Thus, given a geometric cross section $\sigma$, the rate of transits through this patch is given by $\Gamma = \langle n_\pbh\sigma v_\pbh\rangle \simeq n_\pbh\sigma v_0$.

We now estimate the cross section $\sigma$ for \emph{detectable} PBH flybys. As a first estimate of the effect on various Solar System objects from a PBH flyby, we consider a simplified picture involving only the Newtonian interaction of two point masses: a PBH of mass $M_\pbh$ and a visible Solar System object of mass $M_\sso$. Since $v_0$ is much larger than typical orbital speeds in the Solar System---e.g., $v_{\mathrm{Moon}} \approx \qty{30}{\kilo\meter/\second}$ and $v_{\mathrm{Mars}} \approx \qty{24}{\kilo\meter/\second}$---we may assume an instantaneous interaction, in which the Solar System object remains at rest while the PBH travels in a straight path along its initial direction $\bb{\hat{v}}$, with an impact parameter $b_\pbh$ with respect to the object. Then the instantaneous force perpendicular to $\bb{\hat{v}}$ exerted on the Solar System object by the PBH flyby will have magnitude
\begin{equation}
    \label{eq:force}
    F_\perp = \frac{GM_\pbh M_\sso}{\bigl(b_\pbh^2 + (v_0t)^2\bigr)}
        \frac{b_\pbh}{\sqrt{ b_\pbh^2 + (v_0t)^2}}
    .
\end{equation}
Since the flyby occurs on a timescale that is very short compared to the orbital period of the Solar System body, we can ignore the effects of other forces on this timescale and compute the imparted momentum from the instantaneous force within the well-known impulse approximation. The instantaneous momentum imparted to the Solar System object is $\du p_\perp = F_\perp\,\du t = F_\perp\dd x / v_0$. The net impulse velocity $\delta v = p_\perp / M_{\sso}$ imparted to the Solar System object by the PBH flyby can then be estimated by integrating this differential impulse over the entire trajectory, yielding
\begin{equation}
    \label{eq:impulse}
    \delta v \simeq
    \int \frac{\du x}{v_0} \frac{ G M_\pbh b_\pbh}{(b_\pbh^2 + x^2)^{3/2}}
    = \frac{ 2G M_\pbh}{b_\pbh v_0}
    .
\end{equation}

We now use \cref{eq:impulse} to estimate the rate of detectable perturbations at the order-of-magnitude level. Suppose that the object is monitored for a time $\Delta t$ following the perturbation. Over this time, the difference induced in the measured distance from Earth, $r$, can be crudely estimated by $\delta r = \delta v\times \Delta t$. The actual value of $\delta r$ is dependent on all of the parameters of the encounter. However, the value $\delta v\times \Delta t$ can be plausibly attained if the encounter accelerates the object along its direction of motion, modifying its orbital period. Over time, the orbit accumulates a slight phase difference with respect to the original trajectory, corresponding to a difference in the distance with respect to Earth.

This shift, $\delta r$, is the difference between the distance $r$ that would be measured with a PBH flyby versus without, as a function of time. In the context of a Solar System model, $\delta r$ would be the \emph{residual} introduced between the model and data when the system is perturbed by a PBH flyby. We will refer to $\delta r$ as the ``residual'' or ``perturbation'' for the remainder of this work. Note that $r$ always denotes the distance between Earth and a given Solar System object. We use $\bb x$ to denote the position of an object in barycentric coordinates, so the distance of an object from the Sun is denoted by $x = |\bb x|$.

For a first estimate of the detection rate, we take a flyby to be detectable if $\delta r$ exceeds the uncertainty $\sigma_r$ of the measured distances, i.e., if $\delta v \times \Delta t \gtrsim \sigma_r$. Saturating the inequality yields the maximal impact parameter for a detectable encounter as
\begin{equation}
    \label{eq:b-max}
    b_{\mathrm{max}}(\Delta t) =
    \frac{2GM_\pbh}{v_0\sigma_r}\times\Delta t
    .
\end{equation}
The impact parameter $b_{\mathrm{max}}$ translates to a cross section $\sigma = \pi b_{\mathrm{max}}^2$ for detectable events. The total observing time (including time before the encounter) is bounded below as $\Delta t_{\mathrm{obs}} \geq \Delta t$, so $N_{\mathrm{obs}} \gtrsim \Gamma\times\Delta t$. Thus, we can solve for the minimum observing time required for the expected number of detectable encounters to exceed 1. We obtain $\Delta t_{\mathrm{min}} = [v_0\sigma_r^2/(4\pi G^2M_\pbh\rho_\pbh)]^{1/3}$, or
\begin{multline}
    \label{eq:t-min}
    \Delta t_{\mathrm{min}}
    \approx
    \qty{26}{\year}
    \left(\frac{M_\pbh}{\qty{e20}{\gram}}\right)^{-1/3}
    \left(\frac{v_0}{\qty{220}{\kilo\meter/\second}}\right)^{1/3}\\
    \times
    \left(\frac{\sigma_r}{\qty{0.1}{\meter}}\right)^{2/3}
    \left(\frac{\rho_\pbh}{\qty{0.4}{\giga\electronvolt/\centi\meter^{3}}}\right)^{-1/3}
    .
\end{multline}
The corresponding value of $b_{\mathrm{max}}$ is given by
\begin{equation}
    b_{\mathrm{max}}(\Delta t_{\mathrm{min}}) = \left(
        \frac2\pi
        \frac{GM_{\pbh}^2}{\rho_\pbh v_0^2}\frac{1}{\sigma_r}
    \right)^{1/3}
    ,
\end{equation}
which is \qty{3.3}{\au} for the benchmark values in \cref{eq:t-min}. This suggests that current and near-future ephemerides are capable of probing PBH DM at masses that are otherwise unconstrained, with sensitivity driven mainly by flybys at distances typical of inner planet orbits. This enormous effective detector area is what enables the ``direct detection'' of such massive DM particles.

Even as crude approximations, these results are only viable in a fairly narrow mass range. At low masses, the result of \cref{eq:t-min} is cut off when $b_{\mathrm{max}}$ becomes comparable to the radius of the object in question. For Mars, this occurs at a very low mass $M_\pbh \sim \qty{e12}{\gram}$, but such a detection would require an implausibly long observing time of order $\Delta t_{\mathrm{min}} \sim \qty{e4}{\year}$. Moreover, such perturbations will be damped by other physical effects on timescales that are long compared to the orbital period of the target object. We anticipate that perturbations could be detectable for PBH masses as low as \qty{e18}{\gram}, but no lower---indeed, our observable breaks down entirely at low masses, since the high number density of PBHs would lead to a high rate of encounters, introducing interference between different flyby signatures and ultimately washing out the signal of interest. At very high PBH masses, when $b_{\mathrm{max}} \gg r$, Earth itself receives a comparable impulse, and $\delta r$ is suppressed. For this reason, we do not expect to constrain PBH masses above \qty{e23}{\gram}. Conveniently, \qty{e23}{\gram} is also the upper bound of the unconstrained mass range for PBH DM.

Note that $\Delta t_{\mathrm{min}}$ has no dependence on the parameters of the target object itself, but only the precision with which the distance between the target and Earth is measured. Naively, this suggests that the Earth--Moon system would be the ideal detector: since the Earth--Moon distance is measured with a precision of $\sigma_r \sim \qty{1}{\milli\meter}$, one might hope to detect objects with $M_\pbh \sim \qty{e17}{\gram}$ on a short timescale of $\Delta t_{\mathrm{min}} \sim \qty{10}{\year}$. This would fully cover the open window for PBH DM. However, as we will discuss in the next subsection, damping effects are more significant in the Earth--Moon system, and the methods we use here are not well suited to estimate $\delta r$ in this case. Instead, we will focus on the distances between Earth and the other inner planets, which guides our estimates in the following sections.

\subsection{Damping of orbital perturbations}
\label{sec:damping}
We emphasize that the estimate in the previous subsection assumes not only that observational measurements of distances can be performed with precision $\sigma_r$, but also that simulations can reproduce these distances with comparable accuracy, so that the residual $\delta r$ can be meaningfully computed. This is no small feat: such simulations are extremely complex and computationally expensive. In this work, we develop only a proof of principle, and we make no effort to reproduce the sophisticated simulation frameworks developed over several decades in the Solar System dynamics community. Instead, we seek a case in which the many physical effects in these simulations should make a small fractional contribution to the \emph{residual} $\delta r$ induced by a PBH flyby, regardless of their impact on the distance $r$ itself.

\begin{figure}\centering
    \includegraphics[width=\columnwidth]{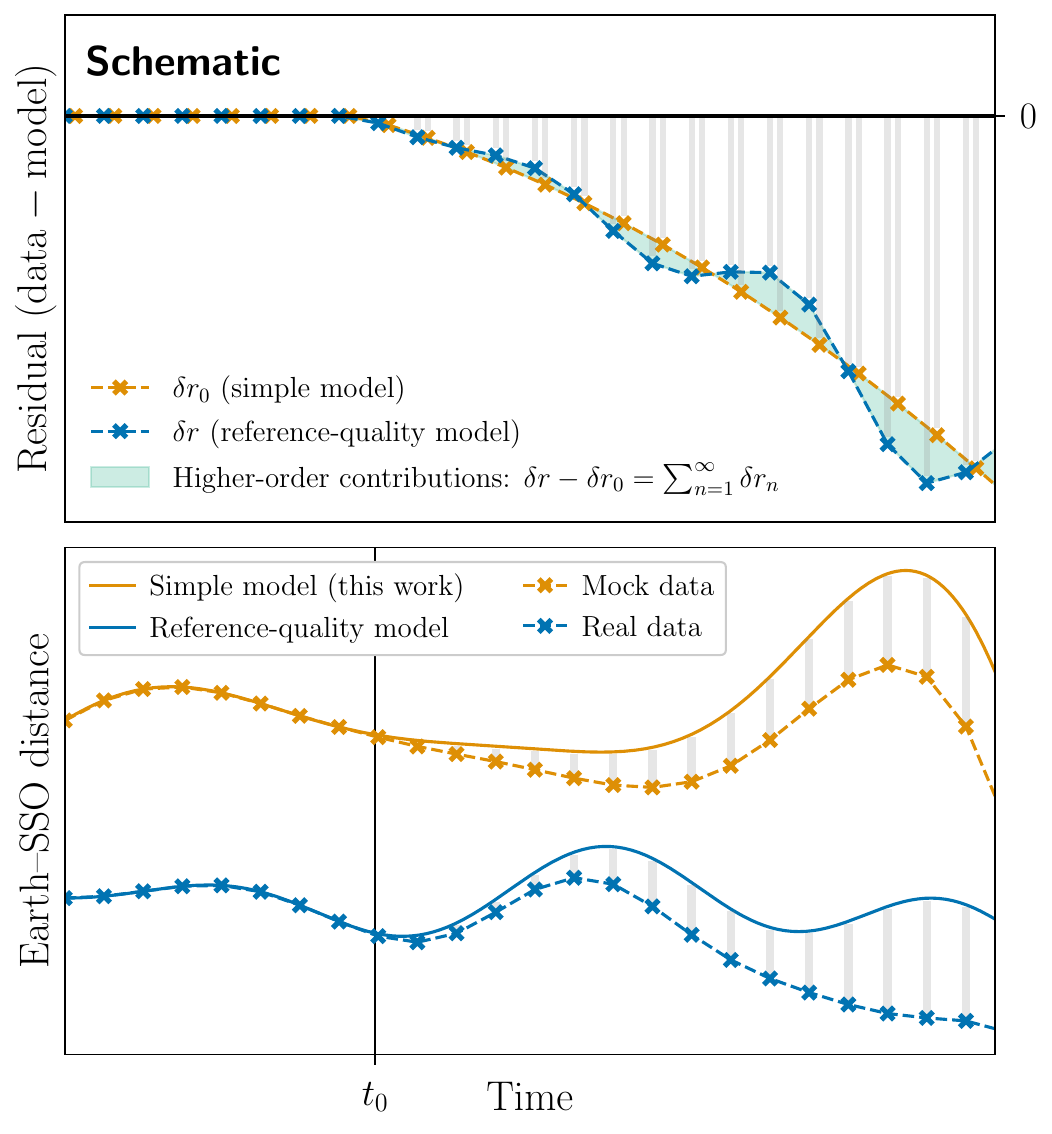}
    \caption{A schematic illustration summarizing the notation and premise of \cref{sec:damping}.
    In this work, we approximate $\delta r$ with a simplified model, which is not sufficient for accurate predictions of real ephemeris data. Still, under certain conditions, we \textit{can} use our simplified model to compute the differences in distance produced by a PBH flyby at time $t_0$.
    The bottom panel shows distances from Earth to a given Solar System object (SSO) as might be produced by the simple model (yellow curve). The simple model closely fits mock data (yellow crosses) before the PBH flyby, but is always a poor fit to actual ephemerides (blue crosses). Fitting these is possible only with a sophisticated, reference-quality model (blue curve).
    Nonetheless, we seek a case in which the residuals between data and model induced by the PBH flyby (gray bars) are comparable in the simple model and the reference-quality model. The top panel shows both sets of residuals, with the difference between them (shaded region) accounted for by terms $\delta r_{n>0}$ in \cref{eq:expansion}.
    We argue in \cref{sec:encounters,sec:simulations} that for some Solar System objects, the higher-order contribution is small compared to the residuals, as illustrated in the top panel. This is precisely why we need not attain the amazing precision of models like DE441 or INPOP21a in order to estimate sensitivity to PBH flybys.}
    \label{fig:schematic}
\end{figure}

More precisely, we estimate $\delta r$ in a perturbative framework, separating the response of the system into an initial perturbation, $\delta r_0$, and higher-order terms $\delta r_n$ describing the backreaction of the system on the \emph{modeled} perturbation due to the \emph{unmodeled} effects. Formally, we write
\begin{equation}
    \label{eq:expansion}
    \delta r = \delta r_0 + \sum_{n=1}^\infty \delta r_n[\delta r_{n-1}]
    ,
\end{equation}
where we note that $\delta r_n$ depends on the functional form of $\delta r_{n-1}(t)$ as a function of time.
We assume that in the presence of the perturbation $\delta r_0$, each unmodeled effect makes an instantaneous contribution $\delta r_1''(t)$ to $\delta r''(t)$, which can be integrated over the observing period to give $\delta r_1(t) = \int_0^t\du {t_1} \int_0^{t_1}\du t_2\,\delta r_1''(t_2)$. 
The perturbation $\delta r_1$ itself sources a correction $\delta r_2$, and so on. We seek a case where the higher-order contribution $\sum_{n=1}^\infty \delta r_n$ is small compared to $\delta r_0$. This framework is illustrated in \cref{fig:schematic}, which shows schematically how $\delta r_0$ can be interpreted as an approximation to $\delta r$.

The interpretation of the higher-order terms is best illustrated by an example. A PBH flyby occurs, and we compute the residual in our simple model: this is the \textit{modeled} initial perturbation $\delta r_0$. However, our model does not include the acceleration due to solar radiation pressure, which is slightly modified in the presence of the perturbation $\delta r_0$. The difference in radiation pressure gives rise to a difference in the acceleration of the object in question, which is integrated to give a correction $\delta r_1^{\mathrm{rad}}$. This correction again modifies the acceleration due to radiation pressure, which gives rise to a correction $\delta r_2^{\mathrm{rad}}$, and so on.

Our computation of $\delta r_0$ from modeled effects is a reliable approximation to $\delta r$ \textit{only} under conditions in which the higher-order terms $\delta r_{n>0}$ can be shown to be small in comparison to the initial perturbation. In what follows, we demonstrate that for the inner planets, $\left|\delta r_1/\delta r_0\right| \ll 1$ over the observational timescale for all effects that are not included in our model but that are included in reference-quality models such as DE441 \cite{JPLeph} and INPOP21a \cite{Pariseph}. We assume, in particular, that $\left|\delta r_1/\delta r_0\right|  \ll 1$ implies that $\left|\delta r_n/\delta r_{n-1}\right|  \ll 1$ for all $n$. This is equivalent to the assumption that our perturbative treatment is valid.

Notice that when $\delta r_0=0$, all of the terms $\delta r_{n>0}$ also vanish, which might lead one to think that the higher-order terms $\delta r_{n>0}$ scale with $\delta r_0/r$. This need not be the case, because these terms may instead scale with other dimensionless ratios $\delta r_0/\ell_{i}$. There are many other length scales $\ell_i$---for example, the distances to other bodies, the radius of the object, the scale of inhomogeneities within the object, and the gravitational parameters $GM_i$ of various bodies. These other length scales can be much smaller than $r$, so that $\delta r_{0}/\ell_i$ can be large even when $\delta r_0/r$ is small.

Using this perturbative language, we now identify specific unmodeled contributions and assess their significance for our computations. The simple estimates of the previous section neglect several physical effects that must be included when simulating the motion of visible objects within our Solar System to the accuracy attained by reference-quality models. In this section, we study the size of the ratio $\left|\delta r_1/\delta r_0\right|$ for each of the effects which are considered in the detailed numerical simulations of \refscite{JPLeph,Pariseph} but which are not included in our simple model, and we identify cases where the corrections are subdominant over the relevant observational time-scale. 

We consider four categories of corrections, in roughly descending order of importance for Solar System flybys:
\begin{enumerate}
    \item\label{it:n-body}
    \emph{Inclusion of additional bodies.}
    State-of-the-art computational models of Solar System ephemerides include many Solar System objects beyond the Sun and planets. (We address this point in \cref{sec:simulations}.)
    \item\label{it:newtonian-finite-size}
    \emph{Newtonian finite-size effects.} Since Solar System objects are not point masses, there are tidal forces between extended bodies, which allow energy loss to heating and deformation. The nonvanishing surface area of extended bodies leads to an additional force from solar radiation pressure.
    \item\label{it:relativistic-point-mass}
    \emph{Relativistic point-mass effects.}
    Post-Newtonian point-mass corrections to the gravitational interactions among $N$ bodies, proportional to both $(v / c)^2$ and to the Newtonian potential $GM / r$, modify the acceleration of each object.
    \item\label{it:relativistic-finite-size}
    \emph{Relativistic finite-size effects.} Finally, in the context of extended bodies, post-Newtonian corrections including the Lense--Thirring effect contribute to spin--orbit coupling in particular.
\end{enumerate}
Each of these effects contributes to $\delta r_1$ and higher terms in \cref{eq:expansion}.

It is tempting to think of a PBH flyby of an orbiting body in the Solar System as a perturbation to an equilibrium system in the same way that we think of perturbations to a classical harmonic oscillator.  However, such a simple picture is inaccurate. Indeed, over very long time-scales, the orbits of Solar System bodies are far from stable \cite{Laskar1,Laskar2,BrownRein}. Furthermore, this approach misses the key difference that while energy-dissipating effects, such as friction, tend to damp the perturbations to a classical harmonic oscillator, the opposite can sometimes be true for dissipative effects in complex Solar System orbits over long time-scales (i.e. many orbital periods) \cite{murray_dermott_2000}. Moreover, dissipative effects in the Solar System are driven by finite-size effects that are suppressed by the small sizes of objects in comparison with their distances. 
Nondissipative effects can also correct $\delta r$, particularly many-body effects that can transfer energy from the body in question and redistribute it amongst other Solar System bodies. But ultimately, to good approximation, bodies in the Solar System conserve their mechanical energy on time-scales of order their orbital periods \cite{murray_dermott_2000}. Thus, if a perturber exchanges energy with one of the bodies in the system, there is no mechanism to quickly restore the system's prior configuration.

That said, the observable that we consider here is an extremely small effect, so even a very small amount of differential dissipation or energy exchange could erase the signature of a PBH transit. Thus, in the remainder of this section, we briefly estimate the significance of each of effects \ref{it:newtonian-finite-size}--\ref{it:relativistic-finite-size} above. Effect \ref{it:n-body} is treated in \cref{sec:simulations}. 

\textbf{Newtonian finite-size effects.} Finite-size effects are famously nonnegligible between planets and their moons. Indeed, for the Earth--Moon system, these effects are significant: given that $R_{\mathrm{Earth}} = \qty{6357}{\kilo\meter}$ and the average Earth--Moon distance is $r_{\textnormal{Earth--Moon}} = \qty{3.84e5}{\kilo\meter}$, the suppression of tidal effects is only of order $R_{\mathrm{Earth}} / r_{\textnormal{Earth--Moon}} \sim \num{e-2}$. On the other hand, finite-size effects are much more strongly suppressed for the Earth--Mars system, with $R_{\mathrm{Earth}} / r_{\textnormal{Earth--Mars}} \sim \num{e-5}$. Therefore, for the remainder of this paper, we focus on quantities such as the Earth--Mars distance rather than on the Earth--Moon system. Treating such well-separated objects as point masses is consistent with the treatment in the reference-quality models DE441 \cite{ParkJPL2021} and INPOP21a \cite{INPOP21a,Fienga:2023ocw}. For similar reasons, we neglect spin-orbit coupling, which remains thoroughly subdominant for the systems and scales of interest here. Fully controlling corrections from tidal dissipation in the Earth--Moon system requires more sophisticated simulations, and is beyond the scope of this work.

The effects of solar radiation pressure, on the other hand, are easy to approximate. The incident radiation power is $P_{\mathrm{rad}} \simeq \qty{}{L_\odot} \times (\pi R_\sso^2) / (4\pi x_\sso^2)$, where $R_\sso$ denotes the radius of the object, $x_\sso$ denotes the distance from the Sun, and $\qty{}{L_\odot} = \qty{1.60e12}{\giga\electronvolt^2}$ is the solar luminosity. If all of the light is reflected, then the force is $F_{\mathrm{rad}} = 2P_{\mathrm{rad}} = \tfrac12\qty{}{L_\odot}(R_\sso/x_\sso)^2$. Thus, following a perturbation $\delta x_0 \sim \delta r_0$ directed away from the Sun, the force changes by at most $\delta F_{\mathrm{rad}} \simeq -\qty{}{L_\odot}(R_\sso/x_\sso)^2\,(\delta r_0/x_\sso)$. The differential displacement over one (circular) orbital period is then bounded from above as
\begin{multline}
    \left|\frac{\delta r_{1}^{\mathrm{rad}}}{\delta r_0}\right| \lesssim
    \frac{2\pi^2 R_\sso^2\qty{}{L_\odot}}
    {GM_\sso\qty{}{M_\odot}}
    \\
    \sim \num{e-12}\times
        \left(\frac{M_\sso}{M_{\mathrm{Mars}}}\right)^{-1}
        \left(\frac{R_\sso}{R_{\mathrm{Mars}}}\right)^{2}
    .
\end{multline}
Thus, regardless of the direction of $\delta r$, the displacement induced by differential radiation pressure is minuscule. For a spherical object with the same density as Mars, achieving $\delta r_{1}^{\mathrm{rad}} \gtrsim \delta r_0$ requires $R_\sso \lesssim \qty{10}{\micro\meter}$, so this effect is not relevant even for the smallest visible objects.

\textbf{Relativistic point-mass effects.}
State-of-the-art ephemeris computations incorporate the leading-order post-Newtonian (PN) corrections to the motion of $N$ point masses moving under their mutual gravitation \cite{ParkJPL2021,INPOP21a,Fienga:2023ocw}, as encapsulated in the Einstein--Infeld--Hoffmann--Droste--Lorentz (EIHDL) equations of motion \cite{Lorentz1937,EIH,Will:2018bme}. We may parameterize the EIHDL corrections by writing the spatial acceleration $\bb a_i$ of the $i$th body (in the barycentric reference frame) in the form $\bb a_i = \bb{\tilde a}_i + \delta\bb a_i$, where $\bb{\tilde a}_i = \sum_{j\neq i} GM_j \, \bb{\hat x}_{ij} / x_{ij}^2$ is the usual Newtonian contribution, and
\begin{widetext}
\begin{multline}
\label{eq:EIHDL}
    \delta \bb{a}_i =
    \sum_{j \neq i} \frac{GM_j\,\bb{\hat x}_{ij}}{x_{ij}^2} \left[
        v_i^2 + 2v_j^2 - 4\,\bb v_i \cdot \bb v_j
        - \frac 32\left({\bb{\hat x}}_{ij} \cdot \bb v_j \right)^2
        - 4\sum_{k\neq i} \frac{GM_k}{x_{ik}}
        - \sum_{k\neq j} \frac{GM_k}{x_{jk}}
        - \frac12 \left( \bb{x}_{ij} \cdot \bb{a}_j \right)
    \right]
    \\
    + \sum_{j\neq i} \frac{GM_j}{x_{ij}^2} \left[
        \left(\bb v_i - \bb v_j\right) \bigl[
            \bb{\hat x}_{ij} \cdot \left( 4 \bb v_i - 3 \bb v_j \right)
        \bigr] + \frac 72 x_{ij}\bb a_j
    \right] .
\end{multline}
\end{widetext}
(Recall we use natural units in which $c = 1$.) Here $\bb{x}_{ij} \equiv \bb{x}_{i}-\bb{x}_{j}$ is the vector that points from the $i$th to the $j$th body, $x_{ij}=\left|\bb{x}_{i}-\bb{x}_{j}\right|$ is its norm, and $\bb{\hat{x}}_{ij}$ is the associated unit vector.
The largest contribution to $\delta r$ from the EIHDL corrections on a Solar System object comes from the post-Newtonian acceleration imparted to the Solar System object by the Sun.
As we did for the case of radiation pressure, we consider a perturbation $\delta x_0 \sim \delta r_0$. The EIHDL acceleration then makes an instantaneous contribution to $\delta r^{\prime\prime}(t)$, which gives an integrated contribution to $\delta r_1$ of order
\begin{equation}
    \left|\frac{\delta r_1^{\mathrm{PN}}}{\delta r_0}\right|
    \simeq \frac52 \frac{G\,\qty{}{M_{\odot}}}{x_\sso}\left(
        \frac{3G\,\qty{}{M_{\odot}}}{x_\sso^3}-\frac{2v_\sso^2}{x_\sso^2}
    \right) \times t^2
    ,
\end{equation}
where $v_\sso$ is taken to be the \textit{average} orbital velocity of the Solar System body and $x_{\sso}$ is the object's distance from the Sun at perihelion. For the Earth--Mars system, we arrive at the conservative bound 
\begin{equation}
    \left|\frac{\delta r_1^{\mathrm{PN}}}{\delta r_0}
    \right|_{\textnormal{Earth--Mars}}
    \lesssim \num{e-4}\times
        \left(\frac{t}{\qty{10}{\year}}\right)^2
    .
\end{equation}
On a 10-year timescale, the effect on the Earth--Venus system is about four times larger, at up to \num{5e-4}, and for the Earth--Mercury system, the correction can be as large as \num{5e-3}. Note that the dependence on observation time is quadratic, so a longer observation time has a pronounced effect on the size of the EIHDL correction. For example, over 100 years, the fractional correction for the Earth--Mercury system reaches \num{5e-1}, which is no longer negligible. However, even on this timescale, the correction to the Earth--Mars distance remains small.

\textbf{Relativistic finite-size effects.} Here we evaluate the impact of the Lense--Thirring effect. In the Lense--Thirring effect, also called the gravitomagnetic or ``frame-dragging'' effect, the rotation of the Sun contributes to the precession angular velocity of a Solar System object by an amount $\delta\Omega$, given by \cite{ParkJPL2021,Fienga:2023ocw}
\begin{equation}
    \label{eq:lense-thirring}
    \delta\bb{\Omega}=
    \frac{G}{x^3_{\sso}}\left(
        -\bb{J}_{\odot} +
        \frac{3\left(\bb{J}_{\odot} \cdot \bb{x}_{\sso}\right)\bb{x}_{\sso}}
             {x^2_{\sso}}
    \right), 
\end{equation}
where $\bb{J}_{\odot}$ is the spin angular momentum of the Sun and $\bb x_{\sso}$ is the instantaneous position of the Solar System object with respect to the Sun. Over a time $t$, the accumulated angular displacement due to the Lense--Thirring angular velocity is given by
    $\delta\theta = \int_{0}^{t}\du t_1\,\delta\Omega(t_1)
    \simeq \delta\Omega \times t$.
Taking the motion of the SSO to be parametrically described by an ellipse with semimajor and semiminor axes $a_\sso$ and $b_\sso$, respectively, the SSO accumulates a spatial displacement $\delta x$ in a time $t$ given by
\begin{equation}
    \delta x(t) = \left|
    \begin{pmatrix}
        a_\sso\cos[\theta(t) + \delta\theta(t)]\\b_\sso\sin[\theta(t) + \delta\theta(t)]
    \end{pmatrix}
        -
    \begin{pmatrix}
        a_\sso\cos\theta(t)\\b_\sso\sin\theta(t)
    \end{pmatrix}
    \right|
    .
\end{equation}
Over any observational timescale of interest, we can be assured that the LT precession angle is small, $\delta\theta \ll 1$, so we have $\delta x(t) \simeq \delta\theta\times\sqrt{(a_\sso\sin\theta)^2+(b_\sso\cos\theta)^2}$.

The perturbation $\delta r_0$ induces a correction $\delta\Omega_1$ to the angular velocity of precession, which induces a correction $\delta\theta_1$ to the integrated angular precession, which finally leads to a correction to the displacement $\delta x_1 = \delta\theta_1\times\sqrt{(a_\sso\sin\theta)^2+(b_\sso\cos\theta)^2} \leq a_\sso\,\delta\theta_1$. This displacement correction approximately bounds the correction to the Earth--SSO distance, i.e., $\delta r_1 \lesssim \delta x_1$. We now make this bound explicit by estimating $\delta\Omega_1$. \Cref{eq:lense-thirring} implies that $\left|\delta\Omega\right| \leq 2G\,\qty{}{J_\odot}/b_\sso^3$. If the perturbation $\delta r_0$ changes the semiminor axis from $b_\sso$ to $b_\sso-\delta r_0$, and the semimajor axis from $a_\sso$ to $a_\sso+\delta r_0$, then the induced change in the Lense--Thirring angular velocity $\delta\Omega_1$ is bounded as
\begin{equation}
    \left|\delta\Omega_1\right| \lesssim 
    \frac{6G\,\qty{}{J_\odot}}{b^4}\times\left|\delta r_0\right|
    .
\end{equation}
Taking the spin angular momentum of the Sun to be $\qty{}{J_\odot} = \qty{1.9e41}{\kilo\gram.\meter^2/\second}$ \cite{2016HPAA....4...33C}, this leads to the bound
\begin{multline}
    \left|\frac{\delta r_1^{\mathrm{LT}}}{\delta r_0}\right| \lesssim
    \frac{6G\,\qty{}{J_\odot}}{b_\sso^3}\frac{a_\sso}{b_\sso}\times t
    \\
    \sim \num{e-11}
    \times\left(\frac{b_\sso}{b_{\mathrm{Mars}}}\right)^{-3}
    \left(\frac{t}{\qty{10}{\year}}\right)
    .
\end{multline}
Clearly, then, the Lense--Thirring contribution to $\delta r_1$ is negligible for the inner planets on the timescale of interest.

The above discussion accounts for all effects of interest except for item \ref{it:n-body} in our original list: the inclusion of other Solar System objects. This is intractable analytically, so in the next section, we turn to simulations to study the correction $\delta r_{1}^{\mathrm{NB}}$ due to a complex $N$-body Solar System.

\section{Solar System simulations}
\label{sec:simulations}
\subsection{Features of flybys}
\begin{figure}\centering
    \includegraphics[width=\columnwidth]{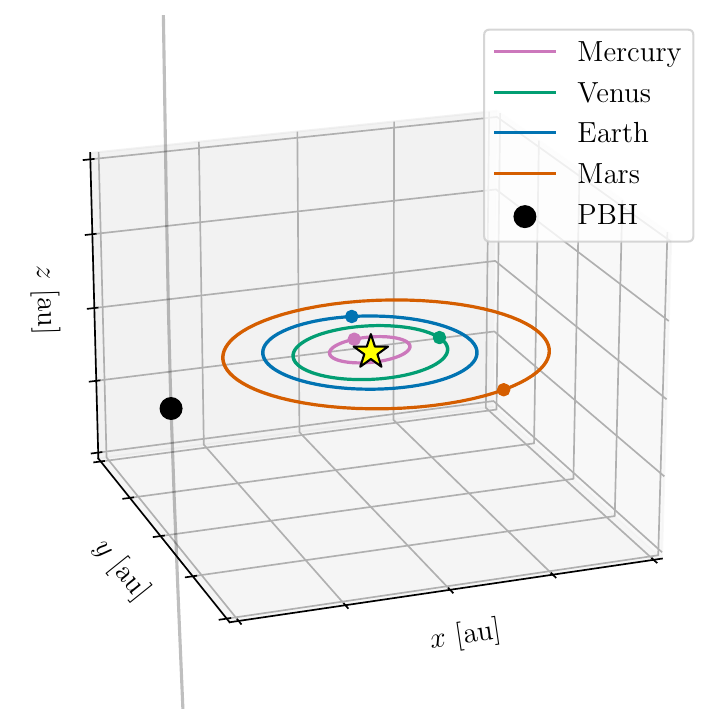}
    \caption{Trajectories for the example PBH encounter considered in \cref{sec:simulations}. Objects beyond the inner planets are omitted from the figure. The PBH perihelion distance is \qty{1.98}{\au}.}
    \label{fig:encounter-orbit-plot}
\end{figure}

We simulate PBH flybys using the \textsc{Rebound} code \cite{Rebound}, taking the masses and initial conditions of Solar System objects (SSOs) from the JPL Horizons database \cite{JPLHorizons,1996DPS....28.2504G}. Our simulations include all of the planets and a selection of smaller objects. To simplify integration, we combine planets and their moons into single particles. The full list of objects is given in \cref{tab:objects}.
We perform the simulations using the WHFast integrator \cite{WHFast}. For high-precision simulations, the IAS15 integrator \cite{IAS15} is usually more appropriate, despite higher computational complexity. We directly compared the behavior of WHFast to IAS15, and found that WHFast is sufficient for the cases considered in this work.

\begin{table}\centering
    {\renewcommand{\arraystretch}{1.2}\setlength{\tabcolsep}{9pt}
    \hrule width \hsize \kern 1mm \hrule width \hsize height 1pt
    \vspace{0.1cm}
    \begin{tabular}{lll}
        \textbf{Object} & \textbf{Mass [\qty{}{\gram}]} & \textbf{Horizons ID}
        \\\hline
        \texttt{Sun} & \num{2.0e+33} & 10 \\
        \hline
        \texttt{Mercury} & \num{3.3e+26} & 199 \\
        \hline
        \texttt{Venus} & \num{4.9e+27} & 299 \\
        \hline
        \texttt{Earth} & \num{6.0e+27} & 399 \\
        ~~~~\texttt{Moon / (Earth)} & \num{7.3e+25} & 301 \\
        \hline
        \texttt{Mars} & \num{6.4e+26} & 499 \\
        ~~~~\texttt{Phobos} & \num{1.1e+19} & 401 \\
        ~~~~\texttt{Deimos} & \num{1.8e+18} & 402 \\
        \hline
        \texttt{Jupiter} & \num{1.9e+30} & 599 \\
        ~~~~\texttt{Io} & \num{8.9e+25} & 501 \\
        ~~~~\texttt{Europa} & \num{4.8e+25} & 502 \\
        ~~~~\texttt{Ganymede} & \num{1.5e+26} & 503 \\
        ~~~~\texttt{Callisto} & \num{1.1e+26} & 504 \\
        \hline
        \texttt{Saturn} & \num{5.7e+29} & 699 \\
        ~~~~\texttt{Titan} & \num{1.3e+26} & 606 \\
        \hline
        \texttt{Uranus} & \num{8.7e+28} & 799 \\
        \hline
        \texttt{Neptune} & \num{1.0e+29} & 899 \\
        \hline
        \texttt{Pluto} & \num{1.5e+25} & 999 \\
        \hline
        \texttt{Ceres} & \num{9.4e+23} & 2000001 \\
        \hline
        \texttt{Vesta} & \num{2.6e+23} & 2000004
    \end{tabular}
    \hrule width \hsize \kern 1mm \hrule width \hsize height 1pt
    }
    \caption{Solar System objects included in the simulations performed in this work. Planets and their moons (indented) are combined into single particles to reduce computational complexity. Names and IDs are drawn directly from the JPL Horizons database \cite{JPLHorizons}.}
    \label{tab:objects}
\end{table}

As discussed in \cref{sec:damping}, our simulations are much simpler than the state-of-the-art simulations used to construct Solar System ephemerides. Relativistic corrections, radiation pressure, and extended-body effects are not included. Additionally, the list of objects is far from exhaustive. Beyond moons, high-quality simulations include at least hundreds of individual asteroids, and additionally include the averaged effects of many smaller asteroids \cite{JPLeph,Pariseph,ParkJPL2021,INPOP21a,Fienga:2023ocw}.

The population of Solar System asteriods is of particular relevance. After all, one might expect that asteroids would have effects on the same order as PBHs in the ``asteroid-mass'' range, which could pose a challenge to the goal of using Solar System dynamics to identify PBHs at these masses. However, the unconstrained window for PBH DM actually extends to masses that are quite high for asteroids. There are very few asteroids in our Solar System with masses of order \qty{e23}{\gram}, and those asteroids are well tracked \cite{BinzelAsteroids}. Moreover, even asteroids at lower masses have manifestly different kinematics: most noticeably, their influence on the ephemerides of other Solar System objects is periodic, as opposed to the single impulse delivered by a PBH. Thus, we expect that the effects of a PBH at the upper end of the ``asteroid-mass'' range can be differentiated from the effects of asteroids.

When adding a PBH to the simulation, we give the PBH a fixed initial speed of \qty{200}{\kilo\meter/\second}, reflecting the typical DM velocity in the Milky Way halo. We specify the flyby with six parameters: one for the mass, three for the initial position, and two for the direction of the initial velocity. The initial position $\bb r_0^\pbh$ is specified with respect to the Solar System barycenter in spherical coordinates $r_0^{\pbh}$, $\Pi_0$, and $\phi_0$. The initial distance $r_0^\pbh$ is an important parameter: it controls the \emph{phase} of the orbits at the time of the encounter. The direction of the initial velocity $\bb v_0^\pbh$ is specified by two angles $\alpha$ and $\beta$ in another spherical coordinate system whose polar axis is aligned with $-\bb r_0^\pbh$. Here $\alpha$ denotes the angle between $-\bb r_0^\pbh$ and $\bb v_0^\pbh$, and $\beta$ denotes the azimuthal angle, chosen such that the projection of a vector with $\beta = 0$ into the $xy$ plane coincides with $\bb{\hat x}$. In this parametrization, $\alpha$ directly corresponds to the impact parameter $b_\pbh$ of the PBH with respect to the barycenter---and given that the encounter is fast, this is approximately equivalent to the perihelion. In particular, for $r_0^\pbh \gg b_\pbh$, we have $b_\pbh \approx \alpha r_0^\pbh$.

We begin with a benchmark simulation taking $M_\pbh = \qty{e21}{\gram}$, $r_0^\pbh = \qty{450}{\au}$, $\Pi_0=\phi_0=0$, $\alpha=2/450$, and $\beta=\pi$. These parameters are chosen so that the impact parameter with the barycenter (and therefore the perihelion) is approximately \qty{2}{\au}. We show a point in time from this simulation in \cref{fig:encounter-orbit-plot}.
The estimates in the preceding sections suggest that an impulse from a PBH flyby should be detectable in Solar System ephemerides. However, in the context of the real Solar System, two critical questions must still be addressed:
\begin{enumerate}
    \item Is a measurable impulse generated at all, given realistic orbital trajectories?
    \item How do complicated $N$-body dynamics affect the evolution of the initial perturbation?
\end{enumerate}
In this section, we answer these questions using $N$-body simulations. We configure simulations with and without a perturbing PBH and directly estimate $\delta r$ (that is, compute $\delta r_0$) by comparing these simulations.

\subsection{Simulation configuration}
We run our simulations for a total of 20 years, and we sample the positions of objects in our simulations with a realistic observational cadence of 20 days (40 days for Mars). At each sample time, we record the distance $r_\sso(t)$ between Earth and each tracked object. We also perform a simulation with no added PBH, obtaining timeseries $r_\sso^{\text{base}}(t)$. For each SSO, we then obtain
\begin{equation}
    \delta r_\sso(t) \approx \delta r_0^{\sso}(t) =
    r_\sso(t) - r_\sso^{\text{base}}(t)
    .
\end{equation}

\begin{figure*}\centering
    \includegraphics[width=0.49\textwidth]{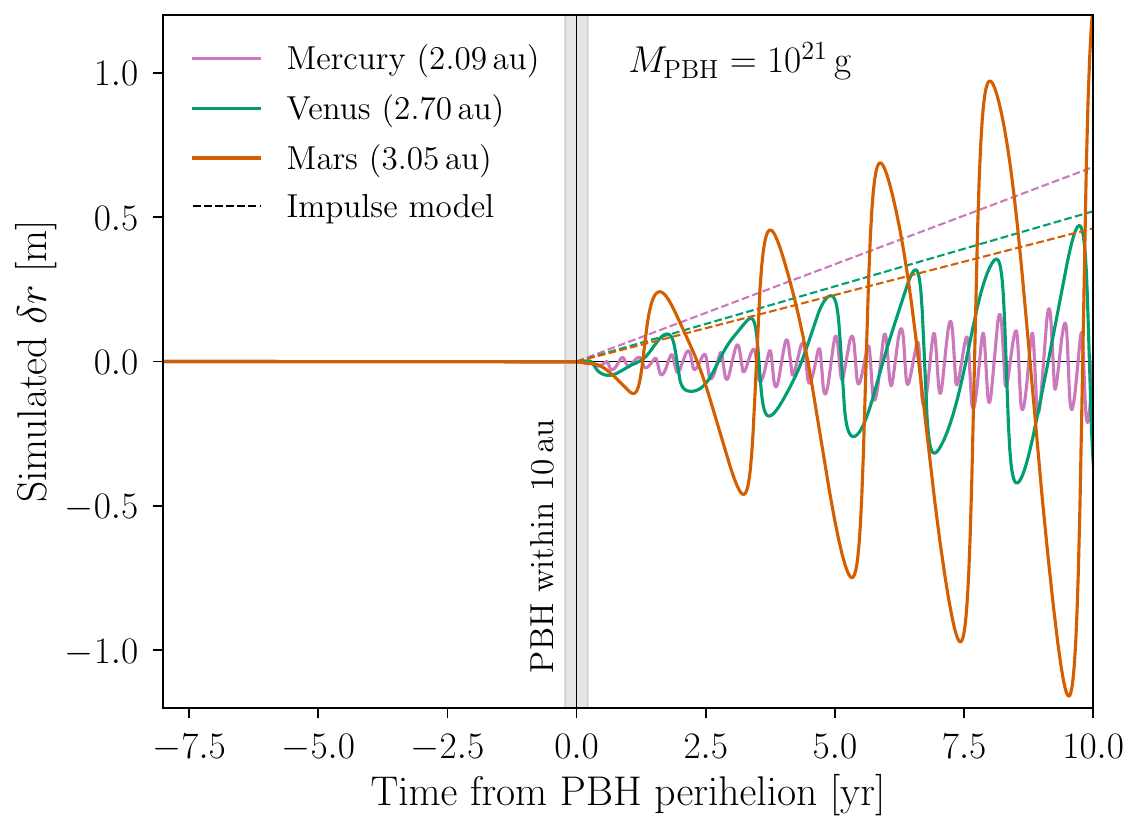}
    \hfill
    \includegraphics[width=0.49\textwidth]{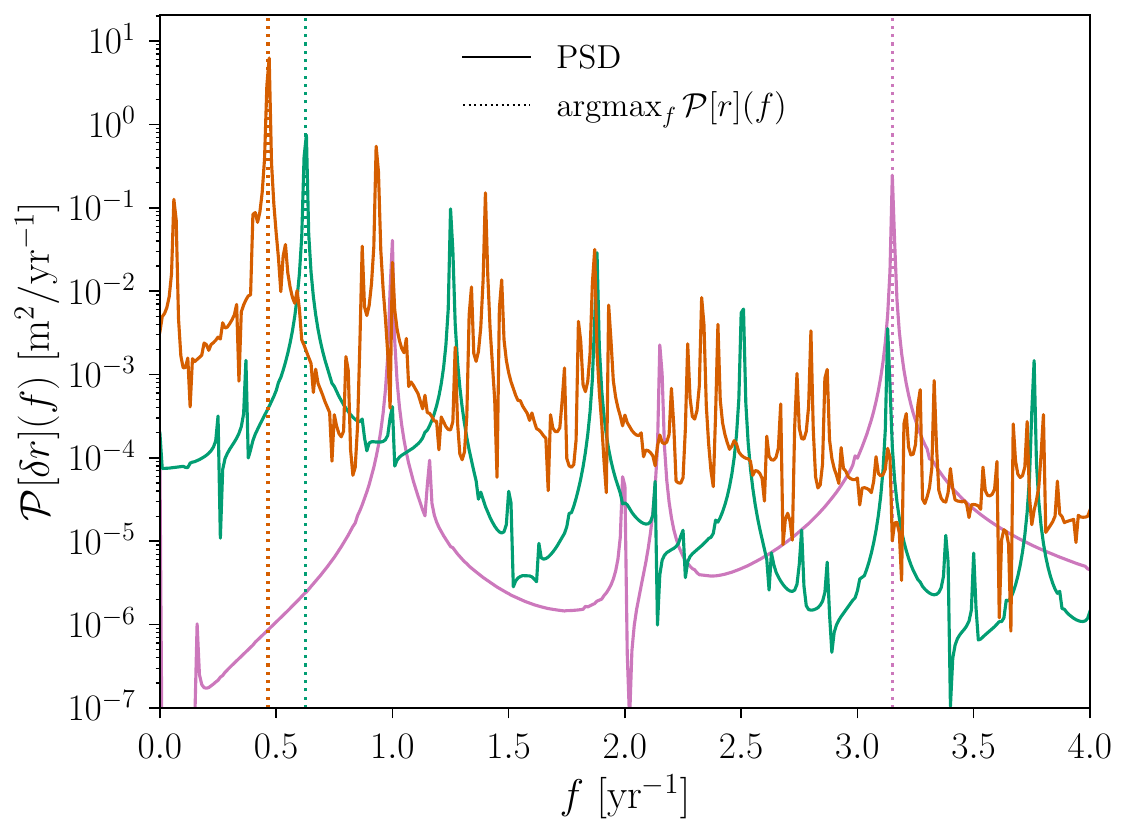}
    \caption{\textbf{Left:} $\delta r$ between Earth and Mercury (purple), between Earth and Venus (green), and between Earth and Mars (orange) for the same encounter shown in \cref{fig:encounter-orbit-plot}. For each object, the distance of closest approach of the PBH is given in parentheses. We set $t=0$ at the PBH perihelion. (Note that the PBH follows a hyperbolic trajectory, so perihelion occurs only once.) The vertical gray band indicates times at which the PBH lies within \qty{10}{\au} of any of the objects shown here. Dashed lines show the simple prediction of \cref{eq:impulse}, i.e., $\delta r = \delta v \times t$. \textbf{Right:} Power spectral density (PSD) $\mathcal P[\delta r]$ for the 100 years following the PBH perihelion. Vertical lines mark the peak frequency of the PSD of the unperturbed distance between Earth and each object, $\mathcal P[r]$. The power peaks sharply at these frequencies, as expected for a slight modification to the orbital parameters.}
    \label{fig:encounter-delta-distance}
\end{figure*}

The value of $\delta r_0$ in this simulation is shown as a function of time for each of the inner planets in the left panel of \cref{fig:encounter-delta-distance}. The simple estimate of the impulse model in \cref{eq:impulse} is shown by the dashed lines. A few qualitative features are immediately clear from the figure:
\begin{enumerate}
    \item The flyby is rapid: the vertical gray band shows the extent of the time period during which the PBH is within \qty{10}{\au} of any of the Solar System objects for which $\delta r$ is shown. The PBH flyby indeed represents a momentary impulse.
    \item For this combination of parameters, $\delta r_0^{\mathrm{Mars}}$ exceeds $\sigma_r = \qty{0.1}{\meter}$ within ${\cal O}(\qty{1}{\year})$ and grows to $\qty{1}{\meter}$ within ten years of PBH perihelion, even though Mars has the largest distance to the PBH among the objects tracked.
    \item The impulse model predicts the size of $\delta r$ at the order-of-magnitude level, but noticeably fails at the $\mathcal O(1)$ level. In particular, the sizes of the impulse predictions follow the opposite order from those of the simulated $\delta r$ values. The difference is especially pronounced for Mercury.
    \item Each perturbation is oscillatory, and indeed nearly monochromatic. The envelope grows almost linearly with time.
\end{enumerate}
We now comment briefly on the implications of each of these features.

The fact that the flyby is rapid justifies the use of the impulse approximation in \cref{eq:impulse}, explaining why this remarkably simple computation succeeds at predicting the rough behavior of $\delta r$. It also establishes a clear distinction between the kinematics of a PBH flyby and what might be expected of a close encounter between two Solar System objects: the PBH flyby is nearly instantaneous on the dynamical timescale of the Solar System. Any successful fit of a PBH transit model to data will sharply identify the time at which the transit took place, whereas Solar System objects interact over longer timescales.

Moreover, one of the objects tracked in this fairly generic simulation exhibits a computed $\delta r$ larger than any of the estimates from the impulse approximation. This serves as the first numerical evidence for the plausibility of our overall premise: that detection of PBHs is possible with current and future data on realistic observational timescales.

However, despite the order-of-magnitude success of \cref{eq:impulse}, the fact that this computation fails to predict the relative magnitudes of the impulses delivered to each of the Solar System objects signifies the importance of the phases of the planets in their orbits and the precise location and direction of the PBH flyby. In particular, this means that \cref{eq:impulse} is not well suited to predict event rates, and we should expect the rates of \cref{eq:t-min} to be substantially modified in the context of simulations. This also means that even an order-of-magnitude estimate of the overall rate requires an \emph{ensemble} of simulations in order to properly marginalize over arrival directions and times.

Finally, we observe that the actual perturbation $\delta r$ is not only oscillatory, but is dominated by a particular frequency: the peak frequency of the variation in the \emph{unperturbed} Earth--SSO distance $r(t)$. Motivated by the nearly-monochromatic oscillations in the left panel of \cref{fig:encounter-delta-distance}, we show in the right panel the power spectral density of $\delta r$ for the 100 years following the PBH perihelion. The dashed vertical lines mark the frequencies $f_r^\sso$ at which the power spectral density of the unperturbed $r(t)$ attains its maximum for each tracked object---essentially, the inverse period of the main component of the distance from Earth to each object. This is to be expected for small changes to the orbital parameters, since the perturbation to the distance will then oscillate on the timescale of the orbital period. The spectrum of $\delta r$ also contains decreasing peaks at the integer harmonics $nf_r^\sso$ for each object. This characteristic profile in frequency space is of great utility, since it enables the use of template-matching techniques to identify PBH flyby events in noisy data. We discuss this possibility further in the coming sections.

At this point, we have performed a simple estimate of $\delta r$ using the impulse approximation in \cref{eq:impulse}; we have argued that all of the neglected physical effects are unimportant to the computation of $\delta r$, apart from the presence of other bodies; and we have performed $N$-body simulations to demonstrate that the signal in $\delta r$ remains viable in a realistic Solar System, and indeed has several valuable properties. Our first set of simulations also suggests that such parameters as arrival direction and the phases of planetary orbits cannot be neglected. Since PBHs would arrive from random directions and at random times, the effects of these parameters should be included into our estimates via an ensemble of simulations. We take up this task in the next section.

\section{Prospects for Detecting PBH dark matter}
\label{sec:detection}
We now combine the results of the previous sections to obtain appropriate figures of merit for potential constraints on PBH dark matter. We frame the expected detection rate $\Gamma$ in language similar to DM direct detection experiments. The detection rate for fixed PBH parameters $\bb\Pi$ is simply the flux $\Phi$ of objects multiplied by the probability $p$ of detection. That is, we have
\begin{equation}
    \label{eq:detection-rate}
    \Gamma = \int\du^n\bb\Pi\,
    \frac{\du\Phi}{\du\bb\Pi}\times p(\bb\Pi)
    .
\end{equation}
The flux of objects is $n\bar\sigma v_0$, where $n$ is the number density, $v_0$ is the typical velocity, and $\bar\sigma$ is the cross section of the ``detector''---in our case, an arbitrary large volume encompassing the inner Solar System. (The larger the volume, the lower the average probability of detection.) We choose this volume to be a sphere of radius $r_{\mathrm{target}} \equiv \qty{50}{\au}$ centered at the Solar System barycenter, so that $\bar\sigma = \pi(\qty{50}{\au})^2$, and we fix $v_0 = \qty{200}{\kilo\meter/\second}$. Then the differential rate is $\du\Gamma/\du\bb\Pi = \bar\sigma v_0\,(\du n/\du\bb\Pi)\,p(\bb\Pi)$.

We estimate $\Gamma$ by taking a simple representative form for the probability of detection: we say that a PBH flyby is detected if the peak value of $\delta r$ exceeds some threshold value in ratio to the width of the noise in the measurement of $r$. Specifically, we define $\fom$ as the maximum value of the sum in quadrature of $q^\sso\equiv \delta r^\sso/\sigma_r^\sso$ over all objects. That is,
\begin{equation}
    \fom \equiv
    \max_t\sqrt{\sum_\sso\left(\delta r^\sso(t)/\sigma_r^\sso\right)^2}
    .
\end{equation}
This is a figure of merit for the size of the perturbation induced by the flyby with a form motivated by statistical considerations, and we set $p(\bb\Pi) = \Theta(\fom - q_0)$ for some fixed $q_0$, where $\Theta$ is the Heaviside step function. With this definition, $p(\bb\Pi)$ is an indicator function on the PBH flyby parameter space that is 1 for detectable flybys and 0 otherwise. We can then perform the integral in \cref{eq:detection-rate} directly by sampling flybys with impact parameter $b_\pbh < \qty{50}{\au}$, i.e., flybys whose trajectories pass through the target region with cross section $\bar\sigma$.

\begin{figure}
    \centering
    \includegraphics[width=\columnwidth]{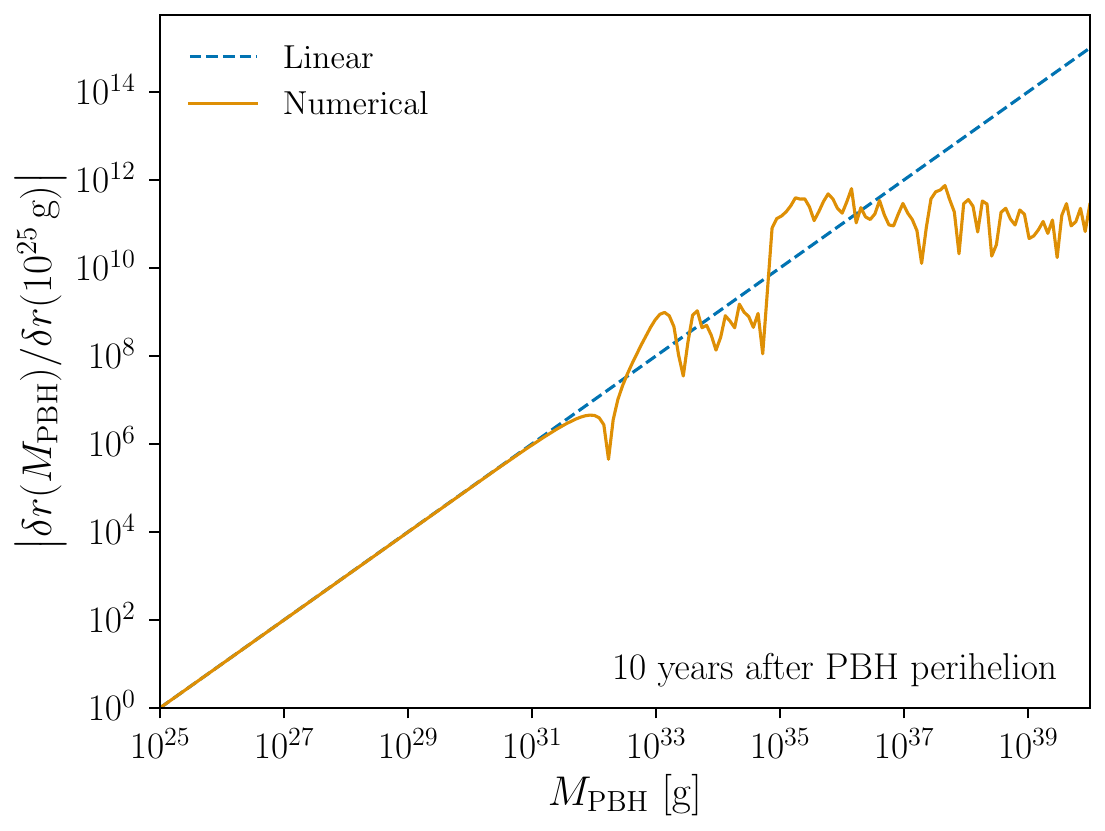}
    \caption{Size of $\delta r$ for Mars as a function of PBH mass. The magnitude is shown at a fixed time, 10 years after PBH perihelion, in ratio to $\delta r$ with $M_{\mathrm{PBH}} = \qty{e25}{\gram}$. All other parameters are consistent with the encounters shown in \cref{fig:encounter-orbit-plot,fig:encounter-delta-distance}. For sufficiently small perturber mass, the size of the perturbation is linear in the mass, as anticipated from \cref{eq:impulse}. For an encounter with these parameters, PBH masses below \qty{e31}{\gram} are in the linear regime.}
    \label{fig:signal-linearity}
\end{figure}

Detailed analysis of the range of realistic values for $q_0$ is beyond the scope of the present work. Obviously, the value of $q_0 = 1$, corresponding to $\max\delta r = \sigma_r$, corresponds to a large effect, and surely should be counted as detectable. However, given the properties of the perturbation in our numerical results, much smaller values of $q_0$ are plausibly accessible. The perturbation $\delta r$ is very long-lived, with a regular and predictable form in Fourier space, which suggests that filtration techniques from signal processing can substantially enhance the sensitivity to small signals. As a reference point, we consider the methods used in the analysis of LIGO data, which routinely enables extraction of signals with an amplitude of order \num{e-4} relative to noise \cite{LIGOSNR}. This level of refinement is \emph{prior} to any analysis of correlations between detectors, an analogue of which is also possible in our case by use of correlations between perturbations to different Solar System objects. Hence, for the remainder of this work, we consider $\num{e-2}<q_0<\num{e-4}$ as a benchmark range.

We numerically approximate the integral of \cref{eq:detection-rate} by sampling values of $\bb\Pi$. We work at fixed mass and speed, so only five parameters remain: $r_0^\pbh$, $\theta_0$, $\phi_0$, $\alpha$, and $\beta$, per the parametrization of \cref{sec:simulations}. We sample uniformly in $\phi_0$ and $\cos\theta_0$, reflecting a random arrival direction. (We neglect the DM ``wind'' associated with the motion of the Solar System in the galactic frame.) We sample $r_0^\pbh$ uniformly on $[\qty{300}{\au},\,\qty{700}{\au}]$, so that all starting points are far from the Solar System, and with perihelion times in a uniform range with a width of $\qty{400}{\au}/(\qty{200}{\kilo\meter/\second}) \approx \qty{9.5}{\year}$. We sample $\beta$ uniformly on $[0, 2\pi]$. We sample $\alpha$ uniformly subject to the restriction that the impact parameter should fall within the cross section of the target, i.e., $b_\pbh = r_0\tan\alpha \leq r_{\mathrm{target}}$. We draw $2^{18}=\num{2.6e5}$ samples from this parameter space using the Sobol sequence method.

Naively, one might expect to evaluate this integral anew for each PBH mass of interest. We simplify our computation by observing that \cref{eq:b-max} predicts that $\delta r$ is linear in $M_\pbh$. \Cref{eq:b-max} is only a simple estimate, but this feature motivates us to check linearity within our numerical simulations: while linearity should not hold for general $M_\pbh$, it should be restored for sufficiently small $M_\pbh$. We thus directly compute $\delta r$ for Mars ten years after the PBH perihelion in our benchmark configuration for a range of different PBH masses, with the results shown in \cref{fig:signal-linearity}. We find that $\delta r$ is indeed linear in the PBH mass as long as $M_\pbh \lesssim \qty{e31}{\gram}$. All of the masses we consider in this work are well within the linear regime, so we can always obtain $\delta r$ at any mass of interest from a simulation at a single mass by simply rescaling the result. We choose to perform all simulations with $M_\pbh = M_\pbh^{\mathrm{base}} \equiv \qty{e27}{\gram}$.

\begin{figure}
    \centering
    \includegraphics[width=\columnwidth]{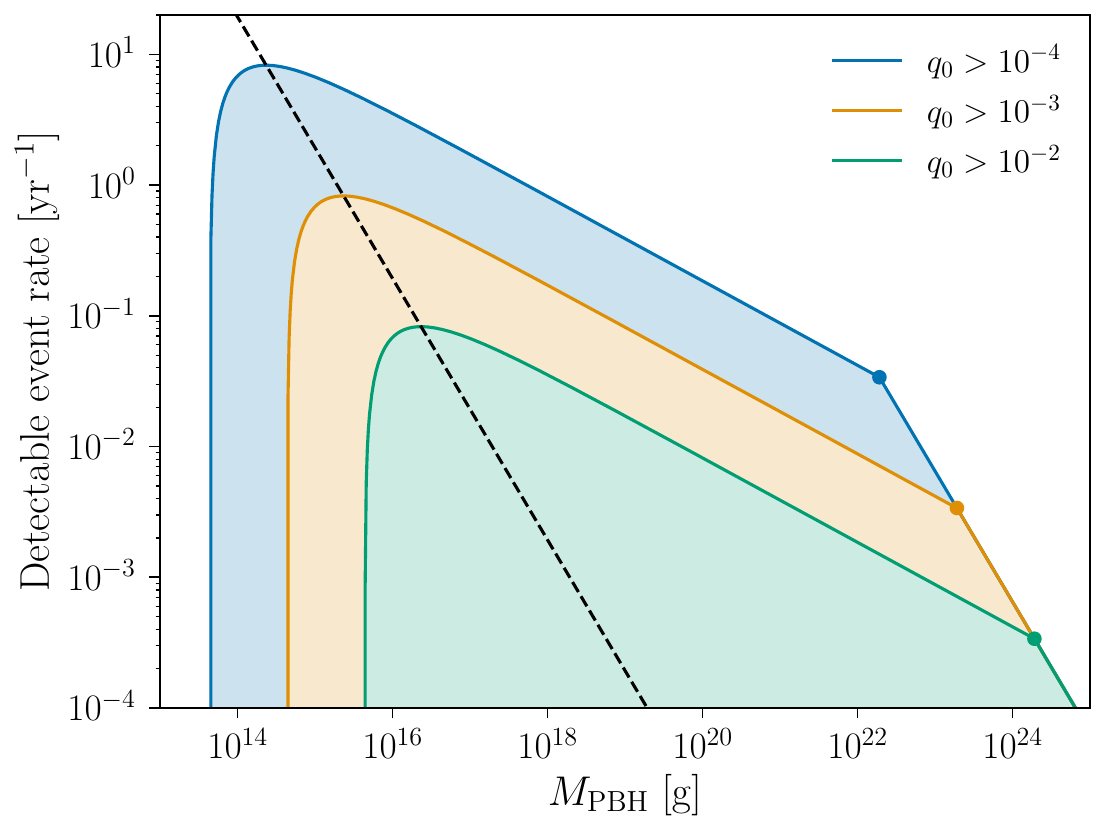}
    \caption{Estimated detectable encounter rate as a function of PBH mass. The rate peaks for PBH masses within the unconstrained asteroid-mass range. Dots indicate the transition from scaling with PBH number density to scaling with flyby detectability ($M_{\mathrm{det}}$). The dashed line shows the peak rate varying the threshold $q_0$, following \cref{eq:peak-rate}. See text for details.}
    \label{fig:encounter-rate}
\end{figure}

The numerical sampling produces a full distribution of $\fom$ at $M_\pbh = M_\pbh^{\mathrm{base}}$. Given that $\fom$ can also be rescaled linearly with the PBH mass, we can readily study the behavior of the detection rate as a function of mass. We fit a power law to the tail of the distribution of $\fom$, truncating at the most extreme values $(\fommin,\,\fommax)$ obtained in our simulations. This simple form for the distribution of $\fom$, combined with the simple form for $p(\bb\Pi)$, leads to a simple and interpretable mass dependence. The survival function of $\fom$ takes the form
\begin{equation}
    \label{eq:survival-function}
    S(\fom) =
    \begin{cases}
        1 & \fom < \fommin \\
        \displaystyle
        \frac{(\fommax)^{\gamma+1} - \fom^{\gamma+1}}
            {(\fommax)^{\gamma+1} -
                (\fommin)^{\gamma+1}}
        & \fommin \leq \fom \leq \fommax \\
        0 & \fom > \fommax,
    \end{cases}
\end{equation}
where $\gamma$ is the index of the power law fit, which we find to be $-1.68$. As noted above, the quantity $S(\fom)$ is computed only for $M_\pbh = M_\pbh^{\mathrm{base}}$. Then the rate with a threshold $q_0$ is estimated as
\begin{equation}
    \label{eq:thresholded-rate}
    \Gamma(q_0) \simeq \frac{\rho_\pbh}{M_\pbh}\times\bar\sigma v_0
        \times S\left(\frac{M_\pbh^{\mathrm{base}}}{M_\pbh}\,q_0\right)
    .
\end{equation}

This estimate is shown for several values of $q_0$ in \cref{fig:encounter-rate}, assuming $\rho_\pbh = \rho_{\mathrm{DM}}$. The features of $\Gamma(M_\pbh)$ are readily understood. At masses above a critical value $M_{\mathrm{det}} = (q_0/\fommin)M_\pbh^{\mathrm{base}}$, the probability of detection is very high, i.e., the survival function in \cref{eq:thresholded-rate} is 1. In this regime, the detection rate is simply the flyby rate, which scales with the PBH number density, and hence inversely with the PBH mass at fixed mass density. The mass $M_{\mathrm{det}}$ is marked with a dot for each curve in \cref{fig:encounter-rate}. As the mass is lowered below $M_{\mathrm{det}}$, the survival function drops below 1, and the detectable event rate is limited by the detection sensitivity. The rate peaks at
\begin{equation}
    \label{eq:peak-rate}
    M_\pbh^{\mathrm{peak}} = M_\pbh^{\mathrm{base}}
        \times(\gamma+2)^{1/(\gamma+1)}
        \times\frac{q_0}{\fommax}
    ,
\end{equation}
which corresponds to $M_\pbh^{\mathrm{peak}} \approx \qty{2.3e18}{\gram}\times q_0$. This is indicated by the dashed black line in \cref{fig:encounter-rate}. At yet lower masses, the survival function drops rapidly to zero, and the detection rate with it.

Under the assumptions above, \cref{fig:encounter-rate} demonstrates the potential for nontrivial constraints on PBH DM in the range $\qty{e18}{\gram} < M_\pbh < \qty{e23}{\gram}$ with the use of $\mathcal O(\qty{30}{\year})$ of ephemeris data. Above \qty{e23}{\gram}, the rate is suppressed by the very small number density of PBHs in the neighborhood of our Solar System. Below \qty{e18}{\gram}, despite the high rate shown in \cref{fig:encounter-rate}, the flux becomes so large that there are many encounters over the observing period, and our estimates are no longer valid. Backgrounds also become much more significant in this regime, as there are numerous Solar System objects with masses of this order, and not all of them are well tracked.

If future observations identify candidate events, it will be important to distinguish rapid transits by PBHs from likely background sources. The most important distinguishing feature will likely be the trajectory of the perturbing object. As emphasized throughout our analysis, typical speeds for transiting PBHs should be $v \approx \qty{200} {\kilo\meter/\second}$. In contrast, among the \num[group-separator={,}]{465778} objects included within the Jet Propulsion Laboratory's Small-Body Database that have come within \qty{3}{\au} of any Solar System planet since 1900, we find $v = 14.2 \pm \qty{7.7}{\kilo\meter/\second}$, with a maximum speed among that set of $v_{\max} = \qty{105.5}{\kilo\meter/\second}$ \cite{JPLsbdb}. In addition, objects bound within the Solar System tend to be on coplanar trajectories, whereas the direction of transiting PBHs should be uncorrelated with the ecliptic plane.

Beyond trajectory information, it is possible that a baryonic perturber in this mass range could be directly identified. Active searches for transient objects throughout the inner Solar System are able to identify very low-mass objects, such as the interstellar object 'Oumuamua, which has a mass of order \qty{e12}{\gram} and spatial dimensions $\sim\!\qty{100}{\meter} \times \qty{30}{\meter} \times \qty{10}{\meter}$ \cite{Seligman1,Seligman2}. The successful identification of one such object does not mean that all others are found, but it does suggest that if a non-exotic perturber with mass \qty{e20}{\gram} were to transit the inner Solar System, there is a reasonable possibility that the object could be directly detected, and correlated with measured perturbations $\delta r$ in the motion of an SSO like Mars. Additional study of likely backgrounds in the asteroid-mass range remains an important area for further research.

Our findings suggest that the method proposed here is complementary to that of \refscite{Bertrand:2023zkl,Cuadrat-Grzybowski:2024uph}, which propose a means of detecting (or constraining) PBHs within the asteroid-mass range using decades of high-precision tracking data from the network of global navigation satellite systems (GNSS) in Earth orbit. Whereas our proposed method is most sensitive to the mass range $\qty{e18}{\gram} < M_\pbh < \qty{e23}{\gram}$, the GNSS data are most effective for $M_\pbh \leq \qty{e17}{\gram}$.

\section{Discussion and conclusions}
\label{sec:discussion}

Primordial black holes (PBHs) remain a compelling candidate for dark matter (DM). To date, a combination of observational and theoretical constraints leaves open a window $\qty{e17}{\gram} \lesssim M_\pbh \lesssim \qty{e23}{\gram}$ within which PBHs could account for the entire DM abundance \cite{Khlopov:2008qy,Carr:2020xqk,Green:2020jor,Villanueva-Domingo:2021spv,Escriva:2022duf}. Probing this mass range---either to yield a quasi-direct detection of a DM candidate or to further constrain PBHs as DM---has proven difficult. In this paper, we describe a new observable for PBHs as DM, which leverages decades of precision tracking of the motions of objects within the inner Solar System. In particular, we demonstrate that the inner Solar System itself could function as a compact-object detector, by exploiting state-of-the-art observing programs and Solar System ephemerides.

In this paper we have identified robust observables from transient flybys of PBHs within several \qty{}{\au} of the Solar System barycenter; estimated the likely effects on the motion of closely tracked Solar System objects from such flybys; and simulated an ensemble of such flybys with which to estimate a realistic detection rate. Given these simulations, much of the presently allowed mass range $\qty{e17}{\gram} \lesssim M_\pbh \lesssim \qty{e23}{\gram}$ could be successfully probed using the types of data and Solar System simulations that are already available.

Our present $N$-body simulations have neglected subdominant effects on objects' motions---such as Newtonian finite-size effects, relativistic point-mass corrections, and relativistic finite-size effects---and our simulations include only a subset of the huge number of Solar System objects that are included in leading ephemerides programs such as the DE441 model maintained by the Jet Propulsion Laboratory \cite{JPLeph,ParkJPL2021} and the INPOP21a model maintained by the Observatoire de Paris \cite{Pariseph,INPOP21a,Fienga:2023ocw}. Nonetheless, by quantifying the relative contributions to Solar System dynamics from the types of objects and dynamics that our simulations neglect, we have argued that our proposed observable provides a viable and realistic possibility for probing the stubborn, remaining mass range within which PBHs could account for all of DM.

To move beyond the estimates analyzed in this paper, several steps will be necessary. First, because the expected signals $\delta r$ from transient PBH flybys are comparable to the present errors $\sigma_r$ with which the distances between Earth and various Solar System objects are tracked, it will be imperative to use more precise Solar System simulations and decades of high-precision observational data.

Second, upon using more accurate models of Solar System dynamics, one may exploit the quasi-periodic nature of the expected signal from a PBH flyby, as shown in \cref{fig:encounter-delta-distance}, to develop a bank of time-series templates with which observational data may be filtered, akin to standard procedures for projects like LIGO \cite{LIGOSNR}. In addition, observable signals could be extracted from real-world noisy data by focusing on \textit{correlations} among the (perturbed) motions of several Solar System objects, such as Mars and Venus, rather than focusing on single objects alone.

By using decades of high-precision tracking data from various Solar System observing programs, combined with expertise already at hand within the Solar System dynamics community, the intriguing possibility that DM consists of PBHs may soon be investigated within our own Earthly neighborhood---adding a novel search strategy to the decades of direct-detection efforts that have been devoted to finding a well-motivated DM candidate.

\begin{acknowledgments}\ignorespaces
    We thank James Battat, Thomas Baumgarte, Agn\`es Fienga, Jared Machtinger, Olivier Minazzoli, Ken Olum, and Clifford Will for helpful discussions. Portions of this work were conducted in MIT's Center for Theoretical Physics and supported in part by the U.~S.~Department of Energy under Contract No.~DE-SC0012567. T.X.T. is also supported by the MIT Undergraduate Research Opportunities Program (UROP). S.R.G. is supported by the NSF Mathematical and Physical Sciences Ascend postdoctoral fellowship under Award No.~2317018. B.V.L. is also supported in part by a Pappalardo Fellowship from the MIT Department of Physics.
\end{acknowledgments}

\bigskip\bigskip

\appendix*
\section{Recovery of perturber parameters}
In this appendix, we demonstrate how the parameters of a perturbing object can be approximately recovered from measured ephemeris data. The main pitfall in this process is the possibility of degeneracies between PBH parameters and SSO parameters, particularly their masses. The masses of SSOs are best determined by the very same models that are fit to ephemerides, so an anomalous contribution to SSO trajectories from a perturber might in principle be absorbed by modification of SSO masses instead.

To address this issue, we perform a numerical experiment as a proof of principle. Beginning with a variant on the simulated benchmark transit considered in \cref{sec:simulations} (\cref{fig:encounter-orbit-plot,fig:encounter-delta-distance}), we attempt to determine the PBH parameters directly from the mock ephemeris data. In particular, we show that the mass of the transiting object can be determined quite accurately.

Even with our simplified Solar System model, numerically exploring the full parameter space is challenging. The parameter space is large: a PBH transit is characterized by the five parameters described in \cref{sec:simulations}, and we also allow a perturbation $\delta M_\sso$ to the mass of each SSO. Ideally, one would be able to marginalize over all $\delta M$ parameters and accurately determine all five PBH parameters in order to allow for follow-up observations. However, this is likely infeasible. Aside from noise, we expect at least one degeneracy in the PBH parameters: under the impulse approximation, the same PBH on a time-reversed trajectory gives very nearly the same perturbation to the SSOs. Moreover, follow-up would need to be rapid: within a single year, a typical PBH would travel $\sim\!\qty{50}{\au}$ from perihelion.

Instead, our immediate goal is to identify the mass of the transiting object. In particular, if the PBH mass can be statistically distinguished from zero while allowing SSO masses to vary, then ephemerides can at least establish that a transit took place, even if the kinematical parameters of the transit are substantially uncertain. Given the relative simplicity of this objective, we use an appropriately simple method: without foreknowledge of the true parameters of the transit, we search for the best-fit parameters, and use the likelihood ratio test to determine the statistical preference for the best-fit point over the unperturbed Solar System.

We searched for the best-fit parameters using simulated annealing (\texttt{scipy.optimize.dual\_annealing}) on a personal computer. To narrow the parameter space, we use the impulse model (\cref{sec:flybys}) to restrict the range of masses we test. From \cref{eq:impulse}, the impulse model predicts contribution to the residual that increases linearly with time over the simulated time period, at a rate $\delta v \simeq 2GM_\pbh/(b_\pbh v_\pbh)$. The PBH parameters $\{\Pi_0,\phi_0,\alpha,\beta\}$ determine the impact parameter $b_\pbh$. Thus, for any choice of the initial angular position and direction of the PBH, the impulse model makes a concrete prediction for $\delta v/M_\pbh$. We estimate $\delta v$ from the mock data by computing the slope of a line from the start of the perturbation to the highest peak in the simulated $\left|\delta r\right|$. In \cref{fig:encounter-delta-distance}, this would correspond to finding the slope of a line going from the beginning of the encounter to the rightmost peak in each solid curve. This estimate of $\delta v$ then translates to an order-of-magnitude estimate $\tilde M_\pbh$ of the PBH mass. We then allow the mass to vary by a factor of 10 in each direction. Concretely, we take the following steps:
\begin{enumerate}
    \item The PBH mass is parametrized as $\log(M_\pbh/M_\odot)$, for $M_\pbh/\tilde M_\pbh(\Pi_0,\phi_0,\alpha,\beta) \in [1/10, 10]$.
    \item We make an affine transformation to map each parameter to the unit interval $[0, 1]$.
    \item For each parameter point $p$ we consider, we perform a simulation to obtain mock residuals. We compare these residuals to the ``true'' residuals obtained from the benchmark simulation to define a loglikelihood $\ell(p)$.
    \item We use simulated annealing in the unit hypercube $[0, 1]\times\dotsb\times[0, 1]$ to find a parameter point $p^{\mathrm{max}}_\pbh$ maximizing the likelihood.
    \item We use simulated annealing a second time to find the optimal parameter point for the null hypothesis, i.e., fixing $M_\pbh = 0$ and allowing $\delta M_\sso$ to vary for each object. We denote this parameter point by $p^{\mathrm{max}}_\sso$.
    \item We test for significance. The likelihood ratio test statistic is given by $\lambda = 2\left[\ell(p^{\mathrm{max}}_\pbh) - \ell(p^{\mathrm{max}}_\sso)\right]$. We take $\lambda$ to be distributed according to a $\chi^2$ distribution with five degrees of freedom, corresponding to the five additional parameters need to describe a PBH transit.
\end{enumerate}

We first performed this test for a transit with $M_\pbh = \qty{e23}{\gram}$, and otherwise identical to the benchmark transit discussed in \cref{sec:simulations} ($M_\pbh = \qty{e21}{\gram}$). The loss landscape is trivial in $\delta M_\sso$, with $\delta M_\sso = 0$ always favored, but very complicated in the transit parameters, with sharp peaks. Despite the numerical complexity, we readily found an optimal parameter point $p^{\mathrm{max}}_\pbh$ with $\lambda = 21.8$. This corresponds to rejecting the null hypothesis of $M_\pbh = 0$ with $p < 0.00057$. Moreover, the PBH mass at this parameter point is \qty{7.6e22}{\gram}, which compares favorably with the input value of \qty{e23}{\gram}. Thus, the inclusion of modifications to SSO masses is clearly not an obstruction to establishing the ocurrence of a transit. While these masses are important to Solar System dynamics, they are not degenerate with the mass of a perturber.

While the mass is easily estimated, the angular parameters are more difficult to recover accurately. The input parameter values for the benchmark case were $(\Pi_0,\phi_0,\alpha,\beta) = (0, 0, 0.00106, 1)\times\pi$, whereas for this particular optimal point, we find the values $(0.9986, 1.3360, 1.1810, 1.7964)\times\pi$. From the value of $\Pi$, the polar angle of the initial position, it is clear that this optimum is close to the nearly-degenerate time-reversed trajectory with the opposite initial position. The correspondence between the velocities is more obvious in Cartesian coordinates. The input parameters correspond to a velocity unit vector
\begin{equation*}
    \bb{\hat v}_0 = \{-0.003333, 0.000000, -0.999994\},
\end{equation*}
while the recovered parameters correspond to
\begin{equation*}
    \bb{\hat v}_0 = \{ -0.005010, 0.003680, 0.999981 \}.
\end{equation*}

One might account for the degenerate optimum by directly testing points in the vicinity of the time-reversed optimum after the initial optimization. For present purposes, we do not attempt such a validation step. Instead, we simply demonstrate that the true optimum can be obtained in the context of a more violent encounter. To that end, we also attempted parameter recovery for a transit with $M_\pbh = \qty{e27}{\gram}$. The transits of such heavy PBHs should not be observable given the local DM density and existing constraints in this mass range, but we nonetheless use this case to check that our computational pipeline can accurately recover kinematical parameters in a case with a higher signal-to-noise ratio.

For this heavy transit, we obtained an optimum with $\lambda = 25.2$ ($p < 0.00013$), and the parameters of the transit were estimated with high accuracy. The PBH mass was found to be \qty{9.5e26}{\gram}, and the angular parameters were found as $(\Pi_0,\phi_0,\alpha,\beta) = (0.0012, 1.232, 0.0020, 5.046)$. It seems at first that the angles $\phi_0$ and $\beta$ are quite discrepant. However, since $\Pi_0$ is small, the difference in $\phi_0$ is inconsequential except in the definition of $\beta$. In fact, the initial angular position is nearly identical, and the initial velocity $\bb v_0$ is quite close to the input as well, with $\bb{\hat v}_0 = \{-0.002760, -0.001323, -0.999995\}$. This corresponds to an impact parameter of $b = \qty{1.71}{\au}$, which compares favorably with the input value of $b = \qty{1.79}{\au}$. Given the simplicity of our approach, these examples suggest that the prospects for PBH parameter recovery with more sophisticated methods are extremely promising.

Finally, we emphasize that these optimizations were performed in less than one day on a single computer, and do not reflect the computational techniques that are already used in fitting models of Solar System ephemerides without the inclusion of perturbers. In future work, we will explore these advanced fitting techniques in combination with fully detailed ephemeris models.

\bibliography{references}

\end{document}